\documentclass{svjour3}

\usepackage{graphicx}        

\begin{document}

\bibliographystyle{plain}

\title{Source separation techniques for characterising cosmic ray transients from neutron monitor networks}

\author{T. Dudok de Wit \and {A. A. Chilingarian} \and \mbox{G. G. Karapetyan}}

\institute{T. Dudok de Wit (corresponding author) \at LPCE, CNRS and University of Orl\'eans, 3A av. de la Recherche Scientifique, 45071 Orl\'eans cedex 2, France,
\email{ddwit@cnrs-orleans.fr}
\and
A. A. Chilingarian \at Yerevan Physics Institute, Cosmic Ray Division, 2 Alikhanyan Brothers St., Yerevan 375036, Armenia, \email{chili@crdlx5.yerphi.am}
\and
G. G. Karapetyan \at Yerevan Physics Institute, Cosmic Ray Division, 2 Alikhanyan Brothers St., Yerevan 375036, Armenia, \email{gkarap@crdlx5.yerphi.am}
}

\date{published in Acta Geophysica, July 2008 -- DOI: 10.2478/s11600-008-0038-6}

\maketitle

\begin{abstract}
The analysis of weak variations in the energetic particle flux, as detected by neutron or muon monitors, can often be considerably improved by analysing data from monitor networks and thereby exploiting the spatial coherence of the flux. We present a statistical framework for carrying out such an analysis and discuss its physical interpretation. Two other applications are also presented: filling data gaps and removing trends. This study focuses on the method and its various uses.

\end{abstract}

%%%%%%%%%%%%%%%%%%%%%%%%%%%%%%%%%%%%%%%%%%%%%%%%%%%%%%%%%%%%%%%%%%%%%%%%%

\section{Introduction}

Many studies have investigated the possibility for monitoring and forecasting geomagnetic storms and solar energetic particle events from the variability of the energetic particle flux on ground (Bieber \textit{et al.}, 2000; Kudela \textit{et al.} 2000, Chilingarian \textit{et al.} 2003, Mavromichalaki \textit{et al.} 2006, Chilingarian and Reymers 2007).  The diagnostic of such solar and heliospheric events, however, requires the investigation of weak variations in the cosmic ray flux. Turning precursory anisotropies into a proxy for space weather has turned out to be a delicate task. The prime reason for this is the complexity of the underlying physical processes, which depend on the orientation and the geometry of the solar wind disturbances, on kinetic effects in the interaction with the approaching shock, and on other effets as well (Dorman 2006).

A second reason is related to the observations. Neutron and muon monitor networks can be used as a single multidirectional spectrograph, which considerably improves the detection of Ground Level Enhancements (GLEs), of anisotropies and of weak transients in the energetic particle flux obtained from global surveys (Krymsky \textit{et al.} 1966, Usoskin \textit{et al.} 1997, Mavromichalaki \textit{et al.} 2004). The analysis of such network data, however, is often hampered by a number of practical difficulties such as data gaps and noise. Most fluctuations in the energetic particle flux, however, are remarkably coherent since they can be observed by several monitors, albeit with different amplitudes and sometimes with some delay. This redundancy is a formidable asset for extracting variations that are often too weak to be detected with one single station but can be extracted from a network of stations. We show how this can be achieved by multivariate statistical analysis, based on the Singular Value Decomposition (SVD) technique (Vaccaro 1991).

%%%%%%%%%%%%%%%%%%%%%%%%%%%%%%%%%%%%%%%%%%%%%%%%%%%%%%%%%%%%%%%%%%%%%%%%%

\section{Extracting coherent signatures from network measurements}

Our main working hypothesis is that the count rate from each detector can be expressed as a superposition of a few common regimes that are observed simultaneously by all stations, but with different amplitudes. This idea is supported by the remarkable similarity between the variations observed at different stations. We thus express the count rate $c(x,t)$ as a linear combination of separable functions in space $x$ and time $t$, hereafter called ``modes''
\begin{equation}
\label{svd1}
c(x,t) = \sum_i f_i(t) g_i(x) \; .
\end{equation}
This decomposition can be carried out in an infinite number of ways, so we add the constraint of orthonormality : $\langle f_i(t) f_j(t) \rangle = \langle g_i(x) g_j(x) \rangle = \delta_{ij}$, where $\delta_{ij}$ is the Kronecker symbol and $\langle . \rangle$ means ensemble averaging. The decomposition then becomes unique
\begin{equation}
\label{svd2}
c(x,t) = \sum_{i=1}^N A_i f_i(t) g_i(x) \; ,
\end{equation}
with $A_i$, the weight of the $i$'th mode, and the number of modes $N$ generally equals the number of stations. This empirical decomposition is directly provided by the SVD  (Chatfield and Collins 1995), which is closely related to principal component analysis. 

The weights are conventionally sorted in decreasing order $A_1 \ge A_2 \ \cdots \ge A_N \ge 0$.  Heavily weighted modes describe features that have a high spatial coherence and thus are observed by several stations. For that reason, the first modes are of prime interest. As we shall see later, they capture both isotropic and anisotropic variations. Detector and statistical noise on the contrary tends to be deferred to the last modes, because they occur locally in time or in space. Since the method is data-adaptive, the modes may differ a little when a different set of stations is selected. It is important that the stations sample different aspects of the dynamics of the flux. 

A recent generalisation of the SVD, called Independent Component Analysis (ICA), regards the decomposition of multivariate data into a linear combination of modes that are not orthonormal, but independent (Hyv\"arinen and Oja 2000). ICA is today often preferred to the SVD when it comes to handling blind source separation problems. Blind source separation refers to the identification of a small number of unknown ``source'' terms (in our case the different contributions to the variability of the energetic particle flux) from an array of measurements, each of which is supposed to be a linear and instantaneous (but unknown) mixture of the sources. Since this problem is heavily underdetermined, constraints are needed. In the case of SVD, the sources are constrained to be uncorrelated. Statistical independence, however, often is a more realistic assumption for disentangling different physical processes. In our case, both the SVD and ICA give very similar results; we shall therefore focus on the SVD, which is less sensitive to non-stationarity and is computationally more convenient.

%%%%%%%%%%%%%%%%%%%%%%%%%%%%%%%%%%%%%%%%%%%%%%%%%%%%%%%%%%%%%%%%%%%%%%%%%

\section{Application to neutron monitor data}

To illustrate our approach we consider one year of hourly pressure-corrected neutron monitor data, obtained from the Solar-Terrestrial Division at IZMIRAN. After discarding all stations that exhibit large data gaps (more than 30\% of the time interval) or experience stability problems, we are left with 43 stations. Small data gaps and spurious values are eliminated using the technique that will be described below.

\begin{figure}
\centering
\includegraphics[width=0.8\textwidth]{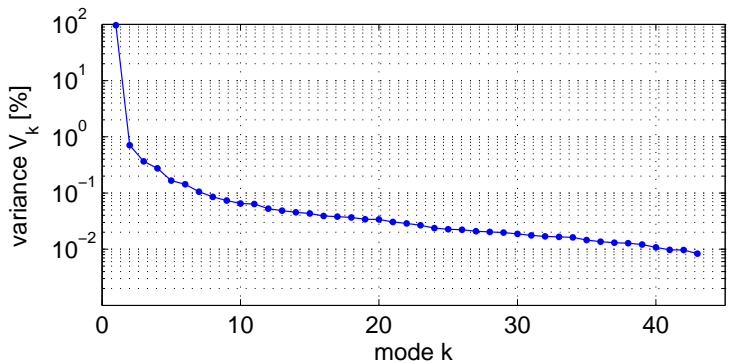}
\caption{Fraction of the variance (in \%) that is explained by each of the 43 modes obtained by SVD.}
\label{fig:gle_weights}       
\end{figure}

The fraction of the variance that is described by the $i$-th mode of the SVD is given by $V_i =  {A_i^2} / {\sum_{j=1}^N A_j^2}$. These values are plotted in Fig.~\ref{fig:gle_weights}, showing a characteristic distribution with a steep falloff, followed by a flat tail. The first modes have the largest variance and capture salient features of the observed variability; three of them describe over $98.2 \%$ of the variance of the data. The modes that are most likely to capture interesting physical signatures are the largest ones (Dudok de Wit 1995), because they describe features that are observed simultaneously by several stations. We expect such features to be the isotropic cosmic ray flux and GLEs. The latter may be very weak, but since they are usually observed coherently by different locations, their actual contribution may become quite significant. In our case, there are about 3 to 5 outstanding modes.

%%%%%%%%%%%%%%%%%%%%%%%%%%%%%%%%% fig
%%%%%%%%%%%%%%%%%%%%%%%%%%%%%%%%%

Since we are interested in variations only of the energetic particle flux, and not in absolute values, for each station we subtract the time average and normalise with respect to the standard deviation. Actually, since the detectors are mostly affected by Poisson noise, one can slightly improve the noise rejection by applying the Anscombe transform prior to the decomposition (Starck and Murtagh 2006). This is equivalent here to taking the square root of the count rate beforehand. We did not apply this transform here.

%%%%%%%%%%%%%%%%%%%%%%%%%%%%%%%%% fig
\begin{figure}
\centering
\includegraphics[width=0.99\textwidth]{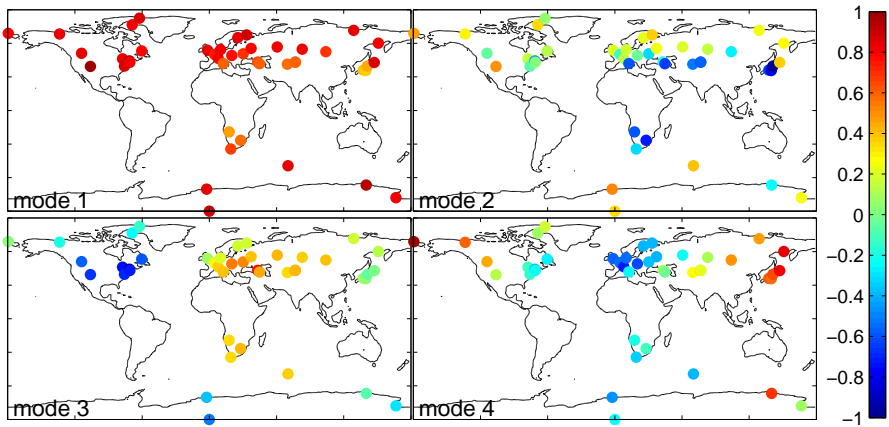}
\caption{Spatial structure of SVD modes 1 to 4. Each dot represents a monitor; the colour expresses the value of the $k$'th spatial mode $g_k(x)$ at that location, normalised to the maximum value of $|g_k(x)|$.}
\label{fig:gle_topos}       
\end{figure}
%%%%%%%%%%%%%%%%%%%%%%%%%%%%%%%%%

The spatial structure of the first 4 modes is displayed in Fig.~\ref{fig:gle_topos}. Several patterns are readily apparent. Mode 1 is a mere weighted average of stations (with weights ranging from 0.94 to 1) whereas mode 2 extracts the difference between low  and high latitude stations, with no longitudinal dependence. Mode 1 should therefore select the large-scale isotropic flux (as discussed by Belov (2000)) whereas mode 2 should express a rigidity dependence. Modes 3 and 4 on the contrary exhibit a clear longitudinal variation, with a mild latitudinal dependence, and so should be related to terrestrial rotation effects or to anisotropies in the flux. Higher order modes in comparison have much less spatial coherence and are more difficult to interpret. Their contribution to the total flux is also weak, so we shall disregard them. Notice that the individual modes can have a negative contribution, but their sum is always positive.

%%%%%%%%%%%%%%%%%%%%%%%%%%%%%%%%% fig
\begin{figure}
\centering
\includegraphics[width=0.99\textwidth]{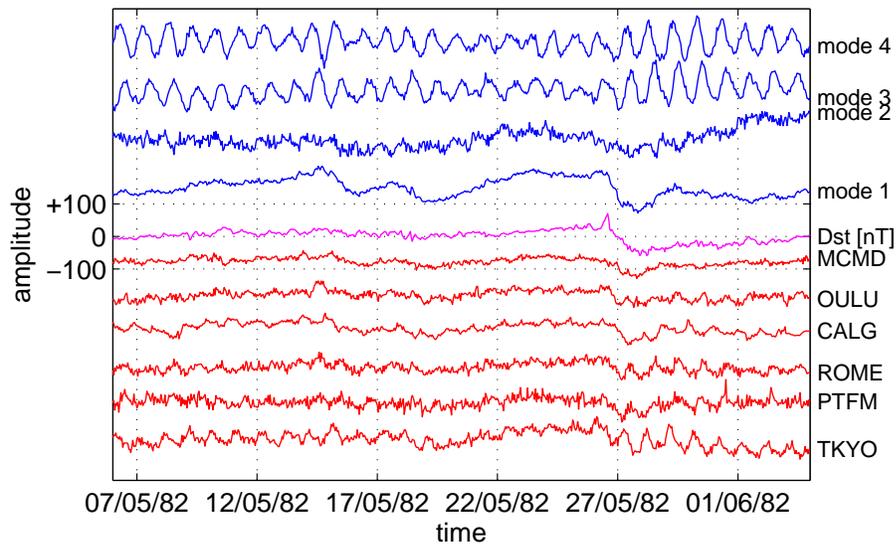} \\
\caption{Excerpt showing, from top to bottom, temporal profiles of modes 1 to 4, the $D_{st}$ index and count rates of 6 stations with increasing rigidity. All but the $D_{st}$ index are plotted with arbitrary units (a.u.). The stations are McMurdo (MCMD), Oulu, Calgary (CALG), Rome, Potchefstroom (PTFM) and Tokyo (TKYO).}
\label{fig:gle_excerpt1}       
\end{figure}
%%%%%%%%%%%%%%%%%%%%%%%%%%%%%%%%%

The physical interpretation of the first SVD modes can be better understood by considering specific events. Figure~\ref{fig:gle_excerpt1} compares the count rates of six stations with different rigidities and the four dominant modes, for a geomagnetically quiet period with a single mild storm. All six stations exhibit similar time evolutions. The most conspicuous signatures are the Forbush decrease associated with the geomagnetic storm that occurs on May 28th, 1982, and the diurnal variations. Their detailed evolution, however, such as the two stages of the Forbush decrease (Cane 2000) is obscured by noise and by local differences between the count rates recorded by each station. 

Mode 1 is remarkably devoid of diurnal variations or short-scale fluctuations. Since this mode if a weighted mean of all stations, with all weights close to unity, we can readily interpret it as a proxy for the isotropic flux. Its advantages over conventional averaging are the following: 
\begin{itemize}
\item the SVD automatically adapts the weighting to the stations in order to extract their common variation. Very similar results are therefore obtained if a different set of stations is used (provided that the stations cover different longitudes and rigidities).
\item Mode 1 is almost  totally devoid of the solar diurnal variation (see below), so that weak variations in the isotropic flux can be accurately monitored. This result is not equivalent to a temporal smoothing of the flux since fast transients are retained.
\item to subtract the isotropic background from the total flux, one simply needs to reconstruct the data from all but the largest mode, i.e. $\hat{c}(x,t) = c(x,t)- A_1 f_1(t) g_1(x)$. 
\end{itemize}

Mode 2 exhibits stronger fluctuations and slowly drifts in time. Fig.~\ref{fig:gle_topos} shows that it is anticorrelated with the rigidity; the correlation coefficient between the two is -0.83. Mode 2 should therefore reflect changes in the spectral hardness of the neutron flux. A decrease in Figure~\ref{fig:gle_excerpt1} implies a relatively stronger contribution from high rigidity stations, which can be interpreted as a harder spectrum. Changes in mode 2 can be ascribed to variations in the primary cosmic ray spectrum, which are typically caused by interplanetary disturbances. Another contribution should come from cutoff rigidity variations, which distort the propagation of cosmic rays and also change the reception coefficients. Such effects are well known to occur during geomagnetic storms (Hofer and Fl\"uckiger, 2000; Belov \textit{et al.} 2005). Notice, however, that mode 2 only captures longitudinally isotropic effects.

Modes 3 and 4 are of a different nature since they are dominated by a 24-hour modulation. These modes are in quadrature; together they describe longitudinally moving patterns (Aubry and Lima 1995) that can readily be associated with the solar diurnal variation.  Notice that the amplitude (defined as $a(t) = \sqrt{f_3^2(t) + f_4^2(t)}$) of the diurnal oscillation drops as expected before the arrival of the interplanetary perturbation on May 28th. 

Let us stress that modes 3 and 4 are not equivalent to bandpass filtered count rates, since they exhibit occasional sharp transients. For that reason, they give a more detailed account of the solar diurnal modulation than standard harmonic (Fourier) analysis. Anisotropies associated with arriving shocks disrupt the regular oscillatory behaviour of these two modes and provide another criterion for detecting anomalous events. Such modes could therefore serve as an input for more elaborate cosmic-ray indices for space weather, as discussed for example by Belov \textit{et al.} (2000), and Kudela and Storini (2006).

%%%%%%%%%%%%%%%%%%%%%%%%%%%%%%%%% fig
\begin{figure}
\centering
\includegraphics[width=0.75\textwidth]{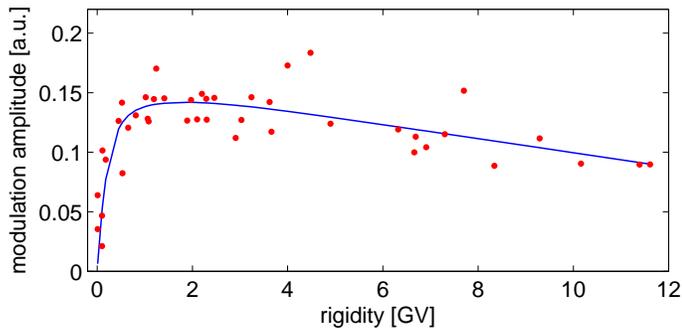} \\
\caption{Modulation amplitude (computed from modes 3 and 4, in arbitrary units) versus the geomagnetic cutoff rigidity.}
\label{fig:gle_modul_rigidity}       
\end{figure}
%%%%%%%%%%%%%%%%%%%%%%%%%%%%%%%%%

Figure~\ref{fig:gle_modul_rigidity} confirms that there is a clear connection between the amplitude of the diurnal modulation and the geomagnetic cutoff rigidity. This connection has received much attention in the literature (see for example Hall \textit{et al.} (1997), and references therein) as it provides insight into the distribution of cosmic rays in the heliosphere. Our main point is that modes 3 and 4 are excellent inputs for carrying out a systematic study of solar diurnal variations.

\begin{figure}
\centering
\includegraphics[width=0.9\textwidth]{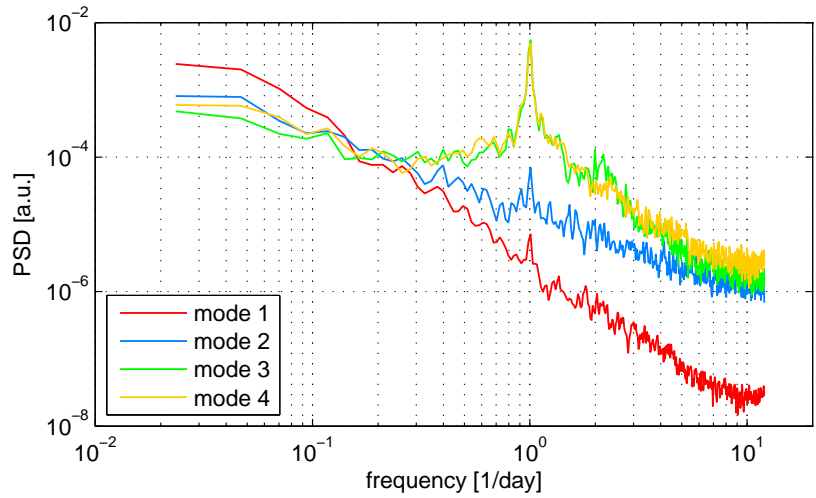} \\
\caption{Power spectral density of temporal modes 1 to 4, estimated using a windowed Fourier transform.}
\label{fig:gle_spect}       
\end{figure}

Interestingly, the power spectral density of temporal modes 1 and 2 reveals a power law scaling over more than two decades, which is almost totally devoid of the 24-hour modulation that dominates in modes 3 and 4, see Fig.~\ref{fig:gle_spect}. The presence of a small residual 24-hour modulation in modes 1 and 2 means that they do not fully succeed in excluding longitudinal anisotropies. This is to be expected from the irregular distribution of the stations. 

The two power laws shown in Fig.~\ref{fig:gle_spect} have distinct spectral indices:  $\alpha_1 = -2.30 \pm 0.03$ for mode 1 and $\alpha_2 = -1.30 \pm 0.03$ for mode 2. The first value is similar to the one found for the $D_{st}$ index and attests a predominance of large-scale transients. Mode 2 in comparison has a much higher level of short-scale fluctuations. These two different spectral indices support the idea that modes 1 and 2 truly capture two different physical processes. 

%%%%%%%%%%%%%%%%%%%%%%%%%%%%%%%%% fig
%%%%%%%%%%%%%%%%%%%%%%%%%%%%%%%%%

\begin{figure}
\centering
\includegraphics[width=0.99\textwidth]{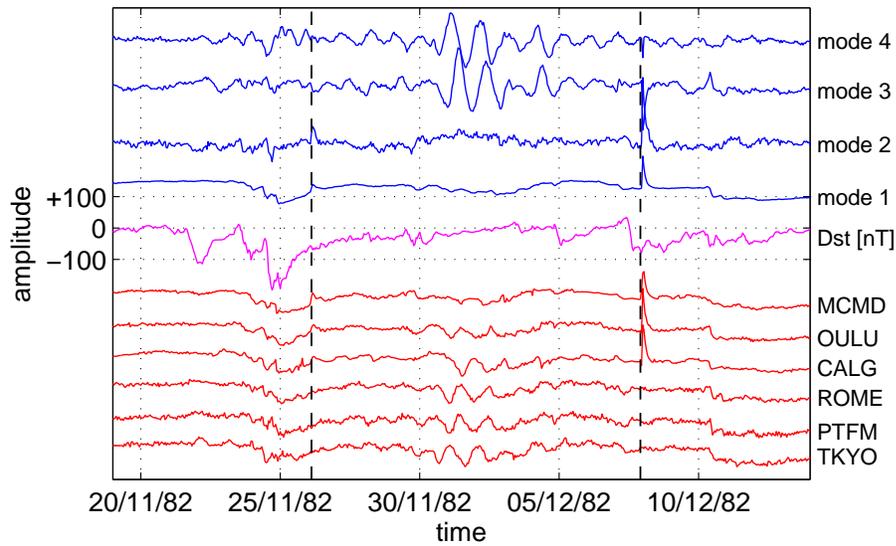}
\caption{Same as Fig.~\ref{fig:gle_excerpt1}, but for a later period. Dashed lines are drawn 2 hours before the onset of the GLEs.}
\label{fig:gle_excerpt2}       
\end{figure}

Figure \ref{fig:gle_excerpt2} shows a second example of a geomagnetically more active period with two GLEs. Both events manifest themselves as a short increase in modes 1 and 2. The increase observed in mode 2 implies that the contribution of GLEs versus that of the cosmic ray background is relatively more important at higher latitudes and can thus be interpreted as a different energy spectrum. Notice in modes 3 and 4 some discontinuities that are not related to diurnal variations but are due instead to the longitudinal anisotropy of the energetic particle flux. The most conspicuous discontinuities occur at the two GLEs. Another one can be observed during the geomagnetic storm of Nov. 24-25, when the propagation of cosmic rays is affected by the changing geomagnetic field.

%%%%%%%%%%%%%%%%%%%%%%%%%%%%%%%%% fig
%%%%%%%%%%%%%%%%%%%%%%%%%%%%%%%%%

%%%%%%%%%%%%%%%%%%%%%%%%%%%%%%%%%%%%%%%%%%%%%%%%%%%%%%%%%%%%%%%%%%%%%%%%%

\section{Filling data gaps}

A second potentially important application is filling data gaps that often plague neutron monitor data. The idea, which is further developed in (Kondrashov and Ghil 2006), is again based  on the redundancy of the observations: because the $M$ most significant modes capture the salient features of the dynamics, one can  replace missing or corrupted samples with values obtained from the truncated expansion $\hat{c}(x,t) = \sum_{i=1}^{M\le N} A_i f_i(t) g_i(x)$. The algorithm has four steps:
\begin{enumerate}
\item Flag all corrupted samples and replace them by some reasonable estimate.
\item Estimate the SVD of the corrected data set.
\item Replace the flagged samples by their value as reconstructed from the most significant SVD modes.
\item Unless the flagged samples have converged to a stable value, proceed to 2.
\end{enumerate}

\begin{figure}
\centering
\includegraphics[width=0.9\textwidth]{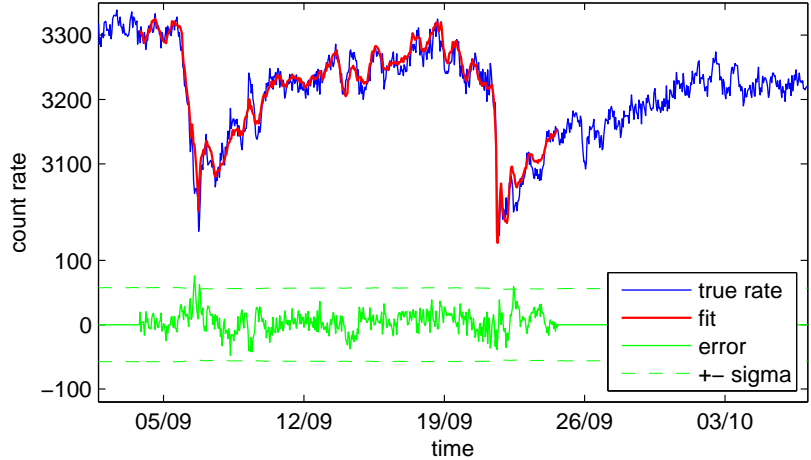}
\caption{Original count rate of the Oulu neutron monitor, its value reconstructed by SVD using $M=4$ dominant modes out of 43, and the residual error, defined as $c(t)-$fit$(t)$. The confidence interval corresponding to $\pm \sqrt{c}$ is also shown.}
\label{fig:gle_oulu}       
\end{figure}

Convergence usually occurs after 5 to 20 iterations. Two advantages of this method are its extreme simplicity (the number of significant modes is the only tunable parameter) and its excellent performance, provided that all the measurements exhibit coherent variations. The result is illustrated in Fig.~\ref{fig:gle_oulu}, in which one month of data from the Oulu monitor were deliberately taken out and subsequently reconstructed by SVD. The good performance of the method is attested by the low variance of the residuals and their almost complete statistical independence.

%%%%%%%%%%%%%%%%%%%%%%%%%%%%%%%%% fig
%%%%%%%%%%%%%%%%%%%%%%%%%%%%%%%%%

%%%%%%%%%%%%%%%%%%%%%%%%%%%%%%%%%%%%%%%%%%%%%%%%%%%%%%%%%%%%%%%%%%%%%%%%%

\section{Trend subtraction}

A third issue is the removal of the slowly drifting trend from a single detector. Trend subtraction, for example, is essential for assessing the statistical significance of weak GLEs. Digital lowpass filters are generally used to extract the slowly varying trend but the design of such filters is a rather delicate and often subjective task. A recent but powerful alternative is Singular Spectrum Analysis (SSA) which is directly related again to the SVD (Elsner and Tsonis 1996). SSA is generally applied to a single time sequence, and the idea is to recover the trend by exploiting its redundancy in time. We thereby assume that the total flux consists of a slowly-varying background level, to which a short transient is added.

So far, we considered spatio-temporal data $\left\{c(x_1,t), c(x_2,t), \ldots, c(x_N,t) \right\}$ and the SVD exploited the coherency in time and in space to extract common modes. We now consider a single station and its $N$ delayed replicates $\left\{c(t), c(t-\tau), \ldots, c(t-(N-1)\tau) \right\}$, where $\tau$ is typically one or a few sampling periods. By applying the SVD to this ensemble, coherency in time only is used to extract different components
\begin{equation}
c(t) = \sum_{i=1}^N B_i h_i(t)
\end{equation}
with $\langle h_i h_j \rangle = \delta_{ij}$ and again $B_1 \ge B_2 \ge \cdots \ge B_N \ge 0$. These components can equally well be obtained by digital filtering, but, interestingly, SSA selects them in such a way that for any $M\le N$, the truncated expansion $\hat{c}(t) = \sum_{i=1}^{M\le N} B_i h_i(t)$ provides the best approximation of the original count rate $c(t)$, in a least-squares sense. For that reason, SSA has been widely advocated for analysing short time sequences and extracting trends in a self-consistent way (Vautard \textit{et al.} 1989).

The two main tunable parameters of SSA are the delay $\tau$, whose default value for short sequences generally is $\tau=1$ sampling period, and the window length $N$, which typically sets the number of samples over which the count rate is averaged. As a rule of thumb, $N/2$ should exceed the duration of the event of interest (the GLE) while being much shorter than the sequence length. 

\begin{figure}
\centering
\includegraphics[width=0.99\textwidth]{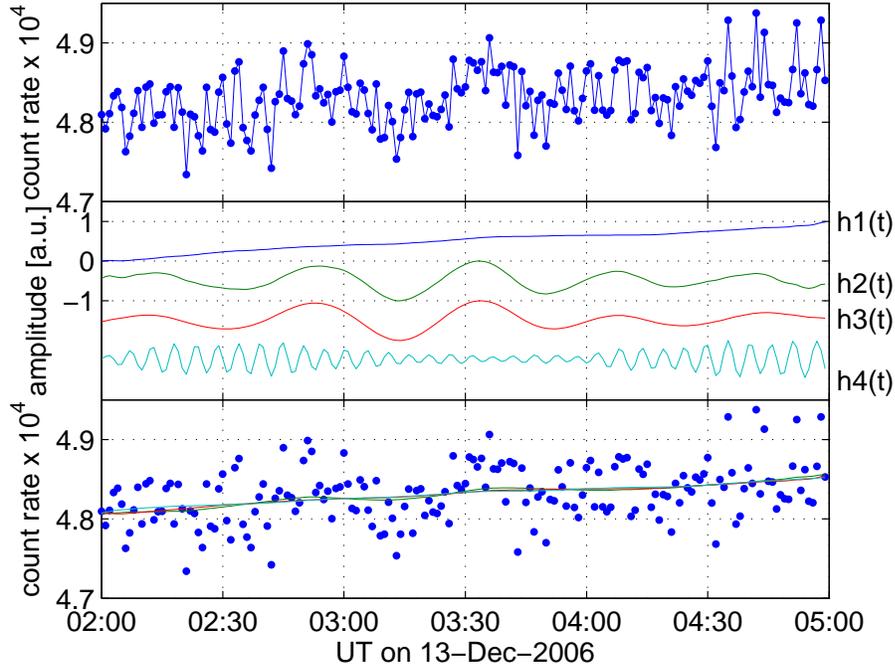}
\caption{From top to bottom: the count rate from the Aragats neutron monitor, the four first components $h_i(t) \ i=1,2,3,4$, estimated for $N=40$, and the trends, compared with the original count rate (dots), for $N=30, 40, 50$.}
\label{fig:gle_trend1}       
\end{figure}

As an example, we consider neutron monitor data measured at Aragats during the night of December 13, 2006, when GLE 70 occurred. The data set consists of 181 samples at a 1-minute cadence. The weak GLE is known to occur at that station between 3:30 and 3:45. The result of the decomposition and the trends are shown in Fig.~\ref{fig:gle_trend1}, for different window lengths. The first component, which captures by far most of the variance, reveals the main trend, while the subsequent ones describe an oscillatory behaviour that is apparent in the raw data. The trend is modelled by taking the first component only. A reasonable value for the window length is $N=30$, but as Fig.~\ref{fig:gle_trend1} shows, $N=20$ or $N=40$ are also acceptable. After removing the trend, we obtain a pure fluctuation count that can be tested statistically. Incidentally, the statistical significance of the GLE signal is highest when the 1-minute count rate is resampled into $t$-minute bins, where $t$ is duration of the signal. In the case of the Aragats neutron monitor, the highest statistical significance, exceeding 4 standard deviations, is obtained with 13-minute bins for the interval running from 3:27 to 3:40.
 
This procedure can be refined in several ways. If the GLE lasts for a long time, as compared to the window length, then its peak may influence the estimation of the trend. One should then flag the samples that belong to the GLE, and estimate the trend without them, in order to get a better estimate of the background.

%%%%%%%%%%%%%%%%%%%%%%%%%%%%%%%%% fig
%%%%%%%%%%%%%%%%%%%%%%%%%%%%%%%%%

%%%%%%%%%%%%%%%%%%%%%%%%%%%%%%%%%%%%%%%%%%%%%%%%%%%%%%%%%%%%%%%%%%%%%%%%%

\section{Conclusions and outlooks}

This study has revealed how small variations in the energetic particle flux, when observed coherently by several neutron monitors, can be significantly enhanced by multivariate statistical analysis based on the SVD method. In particular, coherent patterns that could hardly be detected by one single monitor, can be extracted from the noise level. Although these modes are purely statistical, they convey a physical interpretation. As such, they are valuable inputs for a more detailed analysis of physical properties such as changes in the anisotropy, weak variations in the cosmic ray flux due to solar wind disturbances, etc. Similar indices for cosmic ray activity have been obtained so far by hand, or by using physical models. It is interesting to note that such indices can be recovered by statistical analysis, mainly by assuming that they are uncorrelated.

Several points are worth mentioning. First, since the SVD is a linear method, statistical hypothesis tests can be carried out on weak events. Second, the method is auto-adaptive, so it will automatically search for coherent modes, regardless of the selection of stations, which should of course cover the ensemble of interest.  When using for example surface and underground detectors with different energy responses, differences in the amplitude of the spatial modes $g_k(x)$ are automatically adjusted by the method, which helps in the ill-conditioned problem of spectrum reconstruction. Finally, the SVD is often used in practice to detect anomalous behaviour (drifting count rates, offset errors). A detector that exhibits anomalous behaviour can readily be identified as this difference will tend to be picked up by one single mode.

A natural extension is the analysis of high resolution (e.g. 1-minute) data, which are important for understanding the physical mechanisms associated with GLEs (see for example Plainaki \textit{et al.}, 2006). At such short time scales, propagation effects become significant and no single mode can be associated anymore with the GLE. Our empirical model (Eq. 1) is not appropriate anymore and so transient events will be captured by several modes that have no immediate physical interpretation. Yet, the capacity of a small set of modes to capture the weak signatures of the GLE remains a clear asset of the method.

\begin{acknowledgements}

This work was supported by COST action 724. Comments on the interpretation of the modes by Erwin Fl\"uckiger and on the manuscript by Ludwig Klein are gratefully acknowledged. We would also like to thank the Solar-Terrestrial Division at IZMIRAN for providing the neutron monitor data ({\small \texttt{ftp://cr0.izmiran.rssi.ru/Cosray!/}}) and two referees for helpful comments.  

\end{acknowledgements}

%%%%%%%%%%%%%%%%%%%%%%%%%%%%%%%%%%%%%%%%%%%%%%%%%%%%%%%%%%%%%%%%%%

\end{document}